# 320g Ionization-Heat Cryogenic Detector for Dark Matter Search in the EDELWEISS Experiment

O. Martineau[1] [*], M. Chapellier[2], G. Chardin[3], M. De Jésus[1], J. Gascon[1], S. Hervé[3], A. Juillard[4], M. Loidl[3], X-F. Navick[3], G. Nollez[5]

[1] *IPN de Lyon-UCBL, IN2P3-CNRS, 4 rue Enrico Fermi, 69622 Villeurbanne Cedex, France*
[2] *CEA, Centre d'Etudes Nucléaires de Saclay, DSM/DRECAM, 91191 Gif-sur-Yvette Cedex, France*
[3] *CEA, Centre d'Etudes Nucléaires de Saclay, DSM/DAPNIA, 91191 Gif-sur-Yvette Cedex, France*
[4] *CSNSM, IN2P3-CNRS, Université Paris XI, bat 108, 91405 Orsay, France*
[5] *Institut d'Astrophysique de Paris, INSU-CNRS, 98 bis Bd Arago, 75014 Paris, France*
[*] *Corresponding author. Fax: +33-472-44-80-04 Email: o.martineau@ipnl.in2p3.fr*

**Abstract.** The EDELWEISS experiment is presently using for direct WIMP detection a 320g heat-and-ionization cryogenic Ge detector operated in a low-background environment in the Laboratoire Souterrain de Modane. This detector presents an increase of more than 4 times the mass of previous detectors. Calibrations of this detector are used to determine its energy resolution and fiducial volume, and to optimize the detector design for the 1kg phase of the EDELWEISS-I experiment. Analysis of the calibrations and characteristics of a first series of 320g-detectors are presented.

## 1-INTRODUCTION

Bolometers with simultaneous measurement of ionization and heat provide presently the best existing performances for WIMP direct detection[1,2]. These detectors present a powerful discrimination between electron recoils produced by the radioactive background and nuclear recoils expected from WIMP interactions. In this paper, we present the calibration results of the first 320g Ge Ionization-heat detector used in EDELWEISS-I. The determination by several methods of the detector fiducial volume is also discussed.

## 2- THE 320G IONIZATION-HEAT DETECTOR

The detector[3] is a Ge cylindrical crystal (70 mm diameter and 20 mm thickness) with Al sputtered electrodes for ionization measurement. The top electrode is divided in a central part and a guard ring, electrically decoupled for radial localization of the charge deposition. The thermal sensor consists of a 4 mm$^3$ Neutron Transmutation Doped (NTD) germanium crystal glued on a sputtered gold pad near the edge of the bottom Al electrode. Thus, the residual radioactivity of the NTD sensor is mostly rejected by the guard electrode. The resistance of the DC-polarized sensor was approximately 3MΩ for a temperature of 27 mK, stabilized to within ± 10 µK.

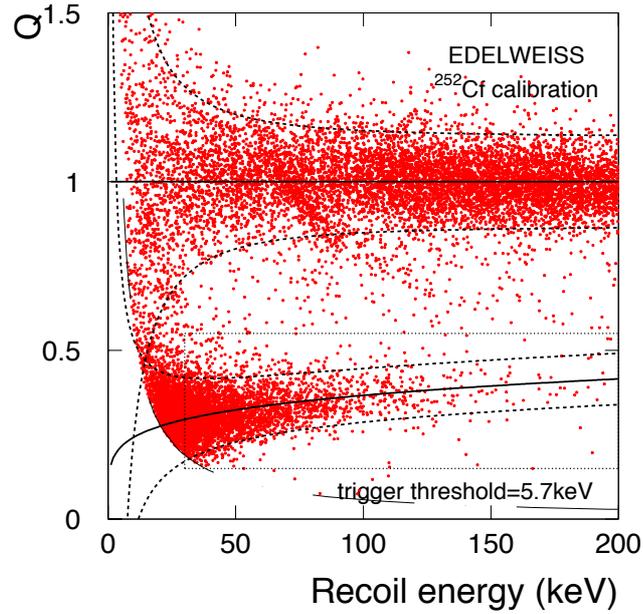

**FIGURE 1.** Distribution of the Q variable as a function of the recoil energy for the $^{252}$Cf data collected in the whole detector volume. The full lines correspond to the average values of Q for electron recoils (Q=1 by construction) and for nuclear recoils (Q=0.16$E_{recoil}^{0.18}$). The two ± 1.645σ bands (90% efficiency) for photons and nuclear recoils are delimited by the thick dotted lines. The dash and dotted line corresponds to the 5.7 keV ionization threshold. The sample selected for the determination of the volume fractions (paragraph 4) is inside the rectangle.

## 3- DETECTOR CALIBRATION

The heat and ionization channels of the detector were calibrated using a $^{57}$Co source. The resolutions measured at 122 keV under +6.37 V polarization were 2.6 keV for the central electrode and 1.8 keV for the guard ring, while the best experimental conditions (NTD polarized at 25 mV and a temperature regulated at 27 mK) gave 3.0 keV at 122 keV for the heat channel. The corresponding baseline resolutions were respectively 1.8 keV, 1.2 keV and 2.2 keV. The ionization resolution was limited by microphonic noise at frequencies varying with time. This also constrained the trigger level of the data acquisition set on the 2 ionization channels. The trigger level was determined with $^{60}$Co runs, the 50% efficiency level being estimated at 5.7 ± 0.3 keV ionization in the best case. 100% efficiency was reached below 8 keV. A recent optimization of the wiring of the ionization channels allowed a significant reduction of the microphonic noise, resulting in a baseline resolution of 700 eV FWHM for the central electrode and a trigger threshold as low as 2.5 keV.

The nuclear recoil zone was calibrated using a $^{252}$Cf source (fig. 1). The average value of the ratio of the ionization energy to the recoil energy, Q, is well described by Q=0.16$E_{recoil}^{0.18}$,(where $E_{recoil}$ is the recoil energy, in keV) a parameterization similar to that obtained in 1997 by EDELWEISS[4] with a 70g detector and by other experiments for this type of detectors[5]. The nuclear recoil zone is determined from the dispersion of the Q variable for neutron interactions and is plotted on fig. 1.

# 4- DETERMINATION OF THE FIDUCIAL VOLUME

Three populations of events are observed in the detector: *center* and *guard* events, corresponding respectively to the exclusive collection of charge on the central and guard electrodes, and a third population called *intermediate events*, when the charge is distributed between the two electrodes. This distribution can be represented using the variable $R=(E_{center}-E_{guard})/E_{total}$ (fig.2), where $E_{center}$ and $E_{guard}$ are the charge signals collected by the central electrode and the guard ring, respectively, and $E_{total}$ the total charge signal. As pointed out elswhere[6,7], the intermediate population may be caused by a macroscopic expansion of the electron-hole cloud, or by multiple scattering interactions inside the detector.

In order to determine the volume fractions corresponding to these three populations, a first method involves electrostatic simulations of the bending of the field lines in the detector. But the exact electrode geometry cannot be easily parameterized, and an alternative method, based on neutron calibrations, has been developed.

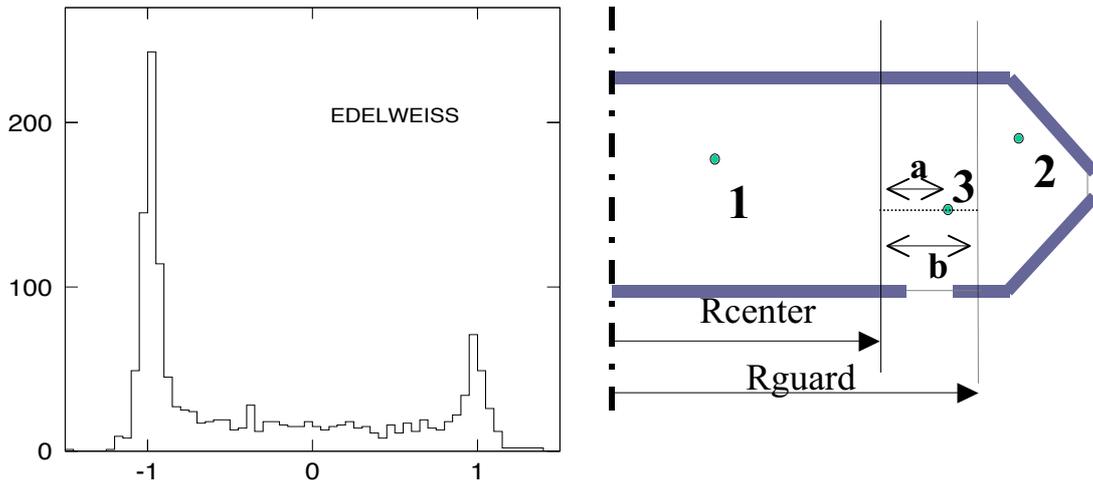

**FIGURE 2A.** Distribution of the R variable for the $^{252}$Cf neutron sample. The center and guard events are peaked around the value R=-1 and R=+1 respectively with a width given by the experimental resolution. The intermediate events are distributed between these 2 values.

**FIGURE 2B.** Distribution of the charge for a simulated interaction with regard to its position. As observed in the experimental case, events are divided into 3 categories: if the event occurs inside an inner zone defined by $R_{center}$ (case 1) or inside an outer ring defined by $R_{guard}$ (case 2), the charge is totally associated to $E_{center}^{simu}$ and $E_{guard}^{simu}$ respectively. This simulates pure center and guard events. When the interaction occurs inside the intermediate volume (case 3), a fraction a/b (a being the distance of the interaction to the inner zone and b the width of the intermediate zone) is attributed to $E_{guard}^{simu}$ and the complementary to $E_{center}^{simu}$. Electrodes are represented by a thick line.

The interactions of neutrons from a $^{252}$Cf source are evenly spread throughout the detector. In a first approximation, the relative proportion of the three populations from a selected sample of $^{252}$Cf neutron events provides a good indication of the relative volumes of the corresponding zones in the detector. A clean sample of such events was obtained by requiring a quenching factor between 0.15 and 0.55 and a recoil energy between 30 and 200 keV (fig. 1). The relative proportion of the three populations in the selected sample was estimated from the R-variable distribution (fig. 2A) to be 41.3 ± 1.0 %, 12.0 ± 0.7 % and 46.5 ± 1.1 % for the center, guard and intermediate events, respectively. Nevertheless, these figures do not reflect exactly the volume fractions of the three zones, since some events are in fact due to multiple scattering of neutrons in the detector. To evaluate the correction arising from this effect, we simulated the response to the $^{252}$Cf source of the detector in its environment with the GECALOR extension of the GEANT code[5]. The detector was then divided into three adjustable zones of cylindrical shape, delimited by two parameters ($R_{center}$ and $R_{guard}$). For each neutron scattering, the simulated recoil energy was converted into two charge signals ($E_{center}^{simu}$ and $E_{guard}^{simu}$), in a relative proportion depending on the position of the interaction (fig. 2B). The experimental data was best described for the values $R_{center} = 2.11$ cm and $R_{guard} = 2.85$ cm. The volume of the inner zone - corresponding in the simulation to an exclusive collection of charge on the central electrode- is thus 46.1 ± 0.9% of the total. The fiducial volume defined elsewhere[1] corresponds to a charge collection on the central electrode greater than 75% of the total. By construction, it is equivalent in this analysis to the cylinder defined by $R_{center} + (R_{center} - R_{guard})/4$, which represents a volume fraction of 54.5 ± 1.5 %. This value can be reproduced by the electrostatic simulations mentioned above to within 5%, which is taken as the systematic error. It is also consistent within statistics to a measurement of the fiducial volume from the 10.4 keV line due to cosmogenic activation.

## 5- DESIGN EVOLUTION AND PERSPECTIVES

Three new 320g detectors will be installed in the LSM to start the data taking in autumn 2001. For these detectors, some large NTD produced by Haller-Beeman have been used as thermistors, the shape of the guard ring has been simplified (in the previous version the electric contact of the central electrode was on the beveled part) and the gold pads on which the electrical contact and the heat link are connected have been made thicker in order to reduce the dislocation production in the single-crystal absorber caused by ultrasonic bonding. This new design should allow us to finalize the determination of the technical specifications before the realization of a first series of 23 detectors for the EDELWEISS-II experiment.

This work has been partially funded by the EEC Network program under contract ERBFMRXCT980167.